# Towards understanding the ordering behavior of hard needles:

# New analytical solutions in one dimension

**Péter Gurin and Szabolcs Varga**

*Institute of Physics and Mechatronics, University of Pannonia, PO Box 158, Veszprém, H-8201*

*Hungary*

# Abstract

We re-examine the ordering behavior of a one-dimensional fluid of freely rotating hard needles, where the centers of mass of the particles are restricted to a line. Analytical equations are obtained for the equation of state, order parameter and orientational correlation functions using the transfer-matrix method if some simplifying assumptions are applied for either the orientational freedom or the contact distance between two needles. The two-state Zwanzig model accounts for the orientational ordering, but it produces unphysical pressure at high densities and there is no orientational correlation. The four-state Zwanzig model gives reasonable results for orientational correlation function, but the pressure is still poorly represented at high densities. In the continuum limit, apart from the orientational correlation length it is managed to reproduce all relevant bulk properties of the hard needles using an approximate formula for the contact distance. The results show that the orientational correlation length diverges at zero and infinite pressures. The high density behavior of the fluid of needles is not resolved.

# 1    Introduction

Orientational ordering is a very interesting property of rod-like particles. In three dimensions hard body models such as the fluid of hard spherocylinders undergoes a first order phase transition from an orientationally disordered (isotropic) phase to an ordered one (nematic) with increasing density. Onsager [1] showed that long ranged order takes place at vanishing packing fraction if the length-to-breadth ratio of the hard body goes to infinity. This make possible to examine the isotropic-nematic (IN) phase transition of infinitely long hard rods exactly using the second-virial theory. Although the real molecules have moderate shape and the IN transition never takes place at vanishing packing fraction, the second virial theory initiated considerable progress in the description of isotropic, nematic and more ordered mesophases [2]. Development of modern density functional theories (DFT) such as the fundamental measure theory is still in the direction to reproduce the exact results of Onsager for infinitely long hard rods [3,4].

In two dimensions (2D), it is widely accepted that there is no true long range orientational order, but a transition of Kosterlitz-Thouless (KT) type [5] takes place between an isotropic phase and a "quasi" nematic phases, where the orientational correlations decay algebraically. This type of transition is shown by Monte Carlo simulations in the fluids of hard needles [6,7], 2D hard spherocylinders [8] and zigzag needles [9,10]. Theories such as the DFT are not able to predict KT-type phase behavior, but they produce second order IN phase transition, where the nematic order is long ranged [11-13]. For example, the virial expansion cannot be truncated at the second term even for the hard needles, where the particle has finite length and zero diameter.

Theoretical examination of one-dimensional (1D) systems has been proved very successful in the last 80 years. If only nearest neighbor interactions are present, exact equation of state and correlation functions can be derived. The famous example is the 1D fluid of hard rods examined by Tonks [14]. Now analytical results are also available for wide classes of fluids with attractive interactions (see [15] and the references therein) and even the pair potential can be obtained analytically from the pair-distribution functions in one dimension [16]. Making the pair potential anisotropic and allowing the particles to rotate freely in two dimensions, it is possible to extend the theoretical descriptions for the orientational ordering properties in the case of nearest neighbor interactions. Several studies are devoted to the system of rod-like hard particles in quasi one dimension such as the rectangles, ellipses and needles [17-25]. The results of these studies are mainly based on the transfer-matrix method, which terminates in an eigenvalue equation. The largest eigenvalue is related to the Gibbs free energy, while the corresponding eigenfunction to the orientational distribution function. For the above mentioned geometries, the eigenvalue problem cannot be solved analytically, because the eigenvalue equation is a complicate integral equation. The numerical calculations show that the fluid of anisotropic hard bodies is orientationally ordered even at very low densities, i.e. the system does not form isotropic phase. The orientational order parameter vanishes at zero pressure (density) and it goes to one for infinite pressure.

Very recently Kantor and Kardar [24] have observed diverging orientational correlation length in the systems of hard needles at the extremely high pressure limit. They have concluded that the diverging correlation length cannot be the indicator of the onset of long range order as the system is ordered. Therefore this behavior is considered as a jamming, where the particles can rotate only in the same rhythm. This conclusion was based on some approximations. Firstly, using a saddle point argument they concluded that the contact distance can be approximated as $\sigma(\varphi,\varphi') \sim |\varphi-\varphi'|$, where $\varphi$ and $\varphi'$ define the orientations of two neighboring needles. On the other hand, instead of solving exactly the model defined by the above approximate contact distance, they used coarse-graining for the angular variables ($\varphi_1, \varphi_2, \ldots$) and analyzed the effective continuum field theory of the approximate model. In our present work we re-examine the phase behaviour of hard needles to check the applicability of the above approximations. To achieve this goal we propose few approximate models for hard needles, which can be solved exactly. We follow two routes. In the first route we impose some restrictions on the orientations of the particles and we examine how the results change if the number of possible orientations is increased. Noteworthy that it has been detected that the 2D system of hard needles with discrete set of orientations undergoes a liquid-gas phase transition [26-28]. Consequently, discretization of the orientation may give rise to unphysical results even for the 1D system of hard needles. To avoid this problem, in the second route we determine analytically the exact thermodynamics of the above mentioned model system proposed by Kantor and Kardar [24], and we compare our results with the prediction of their field theory. We also examine that how close our analytically solvable models are to the system of the hard needles.

Our paper is organized as follows: the transfer matrix method is reviewed shortly in the next section. Working expressions are given for the order parameter and the orientational correlation function. The results of the transfer matrix method for the order parameter, equation of state and orientational correlation function are presented for the approximate models in Section 3. The resulting analytical equations are analyzed in Section 4. Conclusions are given in Section 5.

## 2      Transfer matrix method of quasi-one-dimensional systems

We consider an isobaric ensemble of $N$ particles in quasi-one-dimensional space. Only nearest neighbor interactions are allowed, the particles's centres are restricted to a line and all particles can rotate freely in a plane (see Fig. 1). The interaction potential between two neighbors is defined by

$$u(x_{i,i+1},\varphi_i,\varphi_{i+1}) = \begin{cases} \infty, & x_{i,i+1} \leq \sigma(\varphi_i,\varphi_{i+1}) \\ 0, & x_{i,i+1} > \sigma(\varphi_i,\varphi_{i+1}) \end{cases}, \quad (1)$$

where $x_{i,i+1} = x_{i+1} - x_i$ is the distance between particles $i$ and $i+1$, $x_i$ and $\varphi_i$ denote the position and the orientation of the particle $i$ and $\sigma$ is the orientation dependent contact distance. In hard body systems Eq. (1) guarantees that overlap cannot occur between the particles. Note that we use the $x_1 \leq x_2 \leq ... \leq x_N$ order. The isobaric partition function (configurational part) of a 1D system is

$$Z_{NPT} = \int dx_1..dx_N \, d\varphi_1...d\varphi_N \, \exp\left(-\beta \sum_{i=1}^{N-1} \left(u(x_{i,i+1},\varphi_i,\varphi_{i+1}) + Px_{i,i+1}\right)\right), \quad (2)$$

where $\beta = \frac{1}{k_B T}$ (*T* being the temperature and $k_B$ the Boltzmann constant), and *P* is the pressure. For hard objects (Eq. (1)), the positional integrations can be performed analytically in the partition function

$$Z_{NPT} = \int d\varphi_1 ... d\varphi_N K(\varphi_1, \varphi_2) K(\varphi_2, \varphi_3) ... K(\varphi_N, \varphi_1), \qquad (3)$$

where $K(\varphi_i, \varphi_{i+1}) = \exp(-\beta P \sigma(\varphi_i, \varphi_{i+1}))/\beta P$ can be considered as an element of an infinite-dimensional matrix ($\hat{K}$). In this formalism the matrix product is defined as $K^2(\varphi_i, \varphi_{i+2}) = \int d\varphi_{i+1} K(\varphi_i, \varphi_{i+1}) K(\varphi_{i+1}, \varphi_{i+2})$. Therefore the partition function can be considered as a trace of the matrix $\hat{K}^N$, that is $Z_{NPT} = \text{Tr}\hat{K}^N = \int d\varphi K^N(\varphi, \varphi)$. Since the result of trace operation is independent of the used basis, the partition function can be evaluated most conveniently in the eigenfunction basis, where the matrix $\hat{K}$ is diagonal. Denoting the eigenvalues of $\hat{K}$ in decreasing order as $\lambda_0 > \lambda_1 > \lambda_2 ...$, Eq. (3) simplifies to $Z_{NPT} = \sum_{k=0}^{\infty} \lambda_k^N$, where the following eigenvalue equation provides the eigenvalues

$$\int d\varphi_1 K(\varphi, \varphi_1) \psi_k(\varphi_1) = \lambda_k \psi_k(\varphi). \qquad (4)$$

In the thermodynamic limit ($N \to \infty$) the largest eigenvalue ($\lambda_0$) has the most significant contribution to the partition function, i.e. the natural thermodynamic function (Gibbs free energy) has the following simple form: $\beta G/N = -\log \lambda_0$. Moreover the largest eigenvalue serves the equation of state through the relation $1/\rho = \partial g/\partial P$, where *g=G/N*. It is interesting to note that the eigenfunction ($\psi_0$) corresponds to the largest eigenvalue is related to the orientational distribution function in such a way that $f(\varphi) = \psi_0^2(\varphi)$. The order parameter, which measures the orientational ordering of the particles, is given by

$$S = \sqrt{\langle \sin(2\varphi) \rangle^2 + \langle \cos(2\varphi) \rangle^2}, \qquad (5)$$

where $\langle\sin(2\varphi)\rangle = \int d\varphi\, f(\varphi)\sin(2\varphi)$ and $\langle\cos(2\varphi)\rangle = \int d\varphi\, f(\varphi)\cos(2\varphi)$. In the disordered phase $S$ is zero, while it is equal to one in the perfectly aligned (nematic) phase. It is also important to measure the orientational correlation function, which account for the propagation of the orientational fluctuations along the line. We define it as $g_2(r) = \langle\cos(2(\varphi(0)-\varphi(r)))\rangle - S^2$. Without going into the details [29], it can be shown that

$$g_2(r) = \sum_{k=1}^{\infty} \left(\frac{\lambda_k}{\lambda_0}\right)^{r\rho} (C_k^2 + S_k^2), \qquad (6)$$

where $C_k = \int d\varphi\, \psi_0(\varphi)\cos(2\varphi)\psi_k(\varphi)$ and $S_k = \int d\varphi\, \psi_0(\varphi)\sin(2\varphi)\psi_k(\varphi)$. To obtain the correlation function, the higher order eigenvalues and the corresponding eigenfunctions are needed.

The standard procedure of the determination of the structural properties at a given pressure is the following: 1) It is necessary to know the contact distance between two objects at given orientations. In the case of geometric objects such as the needle, rectangle and ellipse, the calculation of $\sigma(\varphi_1,\varphi_2)$ is just a geometrical problem. Since the molecules are not really hard objects, it is generally accepted that $\sigma(\varphi_1,\varphi_2)$ can be defined without having geometrical model for the molecule. In this way $\sigma(\varphi_1,\varphi_2)$ can be arbitrary function of the orientations for the price of having molecular shape and packing fraction. The only constraint for it is to be consistent with our physical expectations. The well-known example for this is the "hard Gaussian overlap" potential, where the particles do not have geometrical shape, but they interact through the orientation dependent contact distance [30]. 2) One has to find the eigenvalues and the eigenfunctions of the system from Eq. (4).

As an application of the above procedure, let us consider the system of hard circles with diameter $D$. In this case the contact distance does not depend on the orientations, i.e. $\sigma = D$, the largest eigenvalue and the corresponding eigenfunction are $\lambda_0 = \exp(-\beta P D)/\beta P$ and $\psi_0 = 1/\sqrt{\pi}$, respectively, if the orientations are allowed in the range $0 \leq \varphi_i \leq \pi$ ($i=1,2$). From the definition of the equation of state ($\beta/\rho = -\partial \log \lambda_0 / \partial P$) it is trivial to prove that $\beta P = \rho/(1-\rho D)$, which is the well-know equation of state of 1D Tonks gas [14]. Apart from this and some other simple cases [31] where the eigenfunction is constant (isotropic phase), one has to resort to numerical procedures to find the eigenvalues and the eigenfunctions. For hard objects (e.g., needle) we have shown that Eq. (4) can be solved by the standard iterative procedure [32]. It is shown that the needles are always orientationally ordered at any pressure and the most favorable orientation (nematic director) is perpendicular to the line, which means that hard needles are in the nematic phase. In the next section we present new analytical results for the equation of state, the order parameter and the orientational correlation function. In addition, the results of the approximate models are compared with those of hard needles.

## 3     Analytical results

To avoid the lengthy numerical calculations we present three analytically solvable models, which resemble more or less the phase behavior of the quasi 1D hard needles. In the first two models we use the exact hard needle contact distances, but we restrict severely the orientations of the needles. In the simplest model only two configurations are allowed, the particle's orientation can be parallel ($\varphi=0$) or perpendicular ($\varphi=\pi/2$) with the line of the center of masses. This is the so-called Zwanzig approximation [28,33] and is referred to as two-state Zwanzig (Z2) model in the followings. In the straightforward extension of the model, $\varphi=0$, $\pi/4$, $\pi/2$ and $3\pi/4$ orientations are allowed, i.e. the particles can occupy four different states (four-state Zwanzig; Z4). The advantage of Z2 (Z4) approximations is that the eigenvalue equation [Eq. (4)] simplifies to 2x2 (4x4) matrix equation, which can be solved easily. In the third model we treat the orientation as continuous variable without using the exact contact distance of needles, instead an approximate "V"-shaped formula is applied to solve Eq. (4) analytically. The benefit of the "V" model is that we treat the orientational correlations correctly.

a)  **Two-state Zwanzig model**

Let us use the label 1 for $\varphi=0$ configuration (horizontal state) and 2 for the $\varphi=\pi/2$ direction (vertical state). The two-state system has a 2x2 symmetric transfer-matrix ($\hat{K}$) with three different excluded distances: $\sigma(1,1)=l$, $\sigma(2,2)=0$ and $\sigma(1,2)=\sigma(2,1)=l/2$. The vertical state ($\varphi=\pi/2$) is the most favorable, because there are no forbidden regions along the line in this configuration, i.e. the system of parallel needles could behave like an ideal gas. As a result the particles tend to align each other in the direction of vertical state. The explicit form of the transfer-matrix of the two-state Zwanzig model is

$$\hat{K} = \frac{l}{P^*}\begin{pmatrix} \exp(-P^*) & \exp(-P^*/2) \\ \exp(-P^*/2) & 1 \end{pmatrix} \qquad (7)$$

where $P^* = \beta P l$ is the dimensionless pressure. The eigenfunction ($\hat{\psi}$) is now a two component vector satisfying the eigenvalue equation: $\hat{K}\hat{\psi} = \lambda\hat{\psi}$. Two eigenvalues come from the eigenvalue equation $\det(\hat{K}-\lambda\hat{1})=0$. The resulting eigenvalues and the corresponding eigenfunctions in decreasing order are the followings:

$$\lambda_0 = \frac{l}{P^*}(1+\exp(-P^*)) \text{ and } \hat{\psi}_0 = \frac{1}{\sqrt{1+\exp(P^*)}}\begin{pmatrix} 1 \\ \exp(P^*/2) \end{pmatrix}, \qquad (8a)$$

$$\lambda_1 = 0 \text{ and } \hat{\psi}_1 = \frac{1}{\sqrt{1+\exp(P^*)}}\begin{pmatrix} -\exp(P^*/2) \\ 1 \end{pmatrix} \qquad (8b)$$

Since the fraction of particles in the state $i$ is equal to $\psi_0(i)^2$, and the particles prefer the vertical state, the order parameter equation (Eq. (5)) reduces to $S = \psi_0(2)^2 - \psi_0(1)^2$. Using Eq. (8a) the order parameter can be written in a short form

$$S = \tanh(P^*/2). \qquad (9)$$

Using the largest eigenvalue ($\lambda_0$), the equation of state can be obtained from $\beta/\rho = -\partial \log \lambda_0/\partial P$. The resulting equation can be written in dimensionless form as

$$\rho^{*-1} = P^{*-1} + 1/2 - S/2 \qquad (10)$$

where $\rho^* = \rho l$. Finally, it can be seen from Eq. (6) that there is no orientational correlations in the two-state Zwanzig model, because the second eigenvalue ($\lambda_1$) is always zero.

### b) Four-state Zwanzig model

In this model we have four different states. The corresponding orientations are given by $\varphi_i = (i-1)\pi/4$ ($i=1,..,4$), i.e. the angle between the neighboring states is $\pi/4$. The system possesses five different excluded distances and the transfer-matrix is now given by

$$\hat{K} = \frac{l}{P^*} \begin{pmatrix} a^2 & a & a & a \\ a & 1 & b & b^2 \\ a & b & 1 & b \\ a & b^2 & b & 1 \end{pmatrix} \tag{11}$$

where $a = \exp(-P^*/2)$, $b = \exp(-P^*/2\sqrt{2})$. The eigenvalue equation is $\sum_{j=1}^{4} K(i,j)\psi(j) = \lambda \psi(i)$, where $\psi(i)$ is $i$th component of the eigenvector $\hat{\psi} = (\psi(1), \psi(2), \psi(3), \psi(4))$. Again we obtain the eigenvalues from $\det(\hat{K} - \lambda \hat{1}) = 0$. It can be shown that the second largest eigenvalue is given by $\lambda_1 = (1-b^2) l / P^*$, while the rest of them ($\lambda_0 > \lambda_1 > \lambda_3$) can be obtained from the following cubic equation

$$\left(\frac{P^*}{l}\right)^3 \lambda^3 - \left(a^2 + b^2 + 2\right)\left(\frac{P^*}{l}\right)^2 \lambda^2 + \left(1 - a^2\right)\left(1 - b^2\right)\frac{P^*}{l} \lambda + 2a^2(1-b)^2 = 0$$

(12)

As the eigenvalues are lengthy expressions, we do not present them here (see [34] for details). The equation of state and the order parameter are obtained from

$$\frac{1}{\rho^*} = -\frac{\partial \log \lambda_0}{\partial P^*}, \tag{13}$$

and

$$S = \sqrt{\left(\psi_0(1)^2 - \psi_0(3)^2\right)^2 + \left(\psi_0(2)^2 - \psi_0(4)^2\right)^2}, \tag{14}$$

where $\psi_0(i)^2$ is the fraction of molecules in the state $i$ ($i=1,..,4$). Note that there are orientational correlations in this model, because the eigenvalues are not zero. We determine $g_2(r)$ from Eq. (6) using the discretized equations for $C_k$ and $S_k$ functions.

**c) V model**

The next step would be the eight-state Zwanzig model (rotation with 22.5°) with eight eigenvalues and eigenfunctions. The eigenvalue problem results in an 8$^{th}$ order polynomial without having analytical expressions for its roots. However this model would be still far from the continuum limit. To overcome the problem, one possible way is to use an approximate contact distance between two needles. A very simple expression is suggested by Kantor and Kardar [24] and also used by Ansari [25] to examine the high pressure behavior of hard needles. In this approximation the contact distance between two needles has the following form

$$\sigma(\varphi,\varphi') = l|\varphi - \varphi'|/\pi . \qquad (15)$$

This equation is valid for parallel and perpendicular configurations, but it does not give the right value for the specific case of $\varphi=\varphi'=0$. We refer for this model as V-model, because the shape of the $\sigma$ surface reminds us for a V-shaped valley. Eq. (15) can be also considered as a new hard body model, too, where we cannot allocate a geometric shape to the particles. The advantage of the V model is that we have managed to solve it analytically, the system is free rotating and the orientational fluctuations are included. The results of the eigenvalue problem [Eq. (4)] for the V model are the followings. The eigenvalues are given by

$$\lambda_i = \frac{2\pi l}{P^{*2} + \kappa_i^2} \qquad (16)$$

where $\lambda_i$ satisfies the following transcendental equation for even ($i=2j$) and odd ($i=2j+1$) indices

$$\kappa_{2j} \tan(\kappa_{2j}/2) = P^* , \qquad (17a)$$

$$-\kappa_{2j+1} / \tan(\kappa_{2j+1}/2) = P^* . \qquad (17b)$$

The eigenvalues are in the decreasing order ($\lambda_0 > \lambda_1 > \lambda_2 \ldots$) if we search the solution in the following interval $i\pi < \kappa_i < (i+1)\pi$. The corresponding eigenfunctions for even and odd indices are given by

$$\psi_{2j}(\varphi) = \sqrt{\frac{2\kappa_{2j}}{\pi(\kappa_{2j} + \sin \kappa_{2j})}} \cos\left(\frac{\kappa_{2j}}{\pi}(\varphi - \pi/2)\right) \quad (18a)$$

$$\psi_{2j+1}(\varphi) = \sqrt{\frac{2\kappa_{2j+1}}{\pi(\kappa_{2j+1} - \sin \kappa_{2j+1})}} \sin\left(\frac{\kappa_{2j+1}}{\pi}(\varphi - \pi/2)\right) \quad (18b)$$

From Eq. (13) it is easy to show that the equation of state can be written in dimensionless form as

$$\rho^{*-1} = \frac{2(P^* + \kappa_0 \kappa_0')}{P^{*2} + \kappa_0^2} \quad (19)$$

where $\kappa_0' = d\kappa_0/dP^* = \{P^*(1/\kappa_0 + 1/\sin(\kappa_0))\}^{-1}$. The order parameter can be determined from Eq. (5) analytically. It depends on the pressure through $\kappa_0$ as follows

$$S = \frac{\kappa_0^2 \sin \kappa_0}{(\pi^2 - \kappa_0^2)(\kappa_0 + \sin \kappa_0)}. \quad (20)$$

Since the eigenfunctions are proportional to cosine and sine functions, the integrals in the orientational correlation function [Eq. (6)] can be performed analytically. However the formula is very complicate and does not give extra information, so we disregard from the presentation of $g_2(r)$ in this work.

## 4    Comparison of the models

Equation of state and the order parameters of Z2, Z4, V and needle models are shown together in Fig. 2. The pressures of Z2, Z4 and V models are analytical [see Eqs. (10), (13) and (19)], while that of needles (N) is obtained by the numerical solution of the eigenvalue problem [32]. Apart from the ideal gas limit, the curves of different systems deviate substantially, but the densities of discrete models (Z2 and Z4) get closer to each other with pressure. The same phenomenon can be observed in the continuous models (V and N). The high pressure analyses of Eqs. (10) and (13) show that both Z2 and Z4 system behaves as an ideal gas at very high pressures $(\beta P = \rho)$. In the case of continuous V model, the high pressure limit of Eq. (19) is $\beta P = 2\rho$, which is almost valid even at $P^*=10$ ($P^* = \beta Pl$). The needles always have slightly higher pressures, but they have probably the same high pressure- limit. Note that the equation of state of 2D needle system fits $\beta P = 2\rho$ equation in the nematic phase [6]. In the case of 1D needles we are not able to give a definite answer with our numerical method, because the discretization of the orientations suppresses the orientational fluctuations at very high pressures. Irrespectively of the number of discrete orientations, only the most favourable orientation ($\sigma = 0$ in this case), is occupied in the limit of $\beta P \to \infty$, i.e. the system becomes ideal gas ($\beta P = \rho$). Using 256 different orientations for the representation of $\psi(\varphi)$ in the interval $0 \leq \varphi \leq \pi$, we have obtained the ideal gas law from $P^* = 1000$. This shows that the numerical procedures are not useful for studying the high pressure limit. Noteworthy that Z4 model reproduces the equation of state of the needles quite accurately up to $P^*=3$, while Z2 is reasonable only to $P^*=1.5$. This shows that the local orientational fluctuations are large at low pressures.

The order parameter curves also support the above statements (Fig. 2). It can be seen that the order parameter is overestimated by both Z2 (Eq. (9)) and Z4 (Eq. 14)) models, but the curves of Z4 model is always closer to the numerical results. The order parameter goes to one very suddenly in Z2 model and only the vertical orientation is occupied. Therefore the particles becomes parallel and the system behaves as an ideal gas from $P^*=7$ in Z2. The order parameter of Z4 model goes to one smoothly with pressure, because the system has three orientational states ($\pi/4$, $\pi/2$ and $3\pi/4$) with zero contact distance allowing smaller fluctuations to be present. Therefore Z4 model does not behave as an ideal gas even at $P^*=10$. The high pressure limiting value of $S$ for needles cannot be answered by the numerical procedure because only those orientational fluctuations are taken into account, which are larger than the grid size of the discretized model. This problem is not present in the V model, where $\varphi$ is continuous variable and we have analytical solution. The order parameter of V model is quite close to that of needles up to $P^*=1$, but it start to deviate substantially with the pressure. Deeper inspection of Eq. (20) reveals that the order parameter of V model goes towards to ½ with increasing pressure. This means that the system never gets into the completely ordered state ($S$=1), but it becomes partially ordered even at infinite pressure. In the case of needles we have observed much stronger tendency for complete ordering. This result contradicts to the expectation of Kantor and Kardar [24], because they predicted that the V and the needle models should behave identically at very high pressures. This can only happen if the order parameter of needles saturates first and then goes down to the $S$=1/2 limit of the V model. This would be unphysical, because the ordering tendency of needles increases with the pressure. Even if it goes to one, it does not mean that the needles behave as an ideal gas, because infinitesimal orientational fluctuations are present giving extra factor in the equation of state, i.e. $\beta P = 2\rho$. We must mention that the field theory predicts incorrectly that the limiting value of the order parameter is one instead of ½ for V model. This suggests that the effective Hamiltonian proposed in [24] is not applicable for these models.

The other important quantity, which characterizes the ordering behaviour of the system, is the orientational correlation function. In Z2 model there is no orientational correlation, because the rotation of a particle from vertical to horizontal direction reduces the free room available for the other particles, i.e. it is not beneficial even for the vertical neighbors to perform the same rotation. The same happens in the other direction of rotation, because the rotation of a particle from horizontal to vertical direction makes extra room for the neighbouring horizontal particles which feel no any propelling force rotate together with the first particle. Therefore there are no collective rotations in Z2 fluid. The situation changes dramatically in Z4 model, because there are three directions with zero contact distance. As a result, 45° rotation of a particle among 45°, 90° and 135° states stimulates the neighbors to do the same rotations to maximize the free space available for the particles. Only the rotations to the horizontal state do not boost the orientational correlation. These results show that the number of discrete orientations has to exceed two in order to have orientational correlations. Eq. (6) shows that the orientational correlation function always decays exponentially at large distances in Z4, V and N models. Writing the asymptotic behavior of the correlation function in the following form: $g_2(r) \sim \exp(-r/\xi)$, it can be seen from Eq. (6) that the correlation length is equal to $\xi = (\rho \ln(\lambda_0/\lambda_1))^{-1}$. The pressure dependence of this length is depicted in Fig. 3. We can see that $\xi$ diverges at vanishing density in all cases. This is not surprising as the orientational correlations indicate the onset of the orientational ordering at zero density. With increasing density (or pressure) the correlations weaken up to $\rho^* \sim 1$, which corresponds to the average nearest neighbor distance equal to the length of the needle. For higher pressures the correlation length increases again in all models. Z4 and V models show diverging correlation length with pressure. For example, it can be derived from Eq. (16) that $\xi \sim 2\beta P l^2 / 3\pi^2$ for V model. The high pressure property of the correlation length can be interpreted as an onset of orientational jamming, where the orientation fluctuations bring about the large bunches of particles to rotate in the same rhythm as the free room available for the rotation of the particles decreases dramatically with the pressure. Regarding the correlation length of the needles (Fig. 3) suggests that the correlation length does not diverge in the system of hard needles, i.e. it behaves differently from that of V model and it shows saturation. As this result is based on the numerical solution of the eigenvalue equation of needles, no definite conclusion can be drawn for the correlations because of the neglected small orientational fluctuations. Therefore it is necessary to go beyond the V model to give the right answer for this pending issue. One possibility is to find new exactly solvable models, which are closer to hard needles, while the other possibility is to define a proper effective Hamiltonian for the field theory calculations. The main conclusion of the above results is that it is very problematic to predict the high density behavior of the needles due to the presence of infinitesimal fluctuations.

## 5   Conclusions

We have proposed two possible paths to study the ordering behavior of hard needles in quasi one dimension. In the first two models (Z2 and Z4) the particles interacts exactly the same way as the needles do, but the number of possible orientations is limited to two or four directions. In the third model (V) the particles are allowed to rotate freely in two dimensions, but the pair interaction does not follow the rules of the hard needles exactly. We have obtained the following important results.

1) The discretization of the orientations has series effect on the equation of state, order parameter and orientational correlation at high pressures. Irrespectively of the number of possible orientations, the particles align parallel in the vertical state ($S\rightarrow 1$) and the system behaves as an ideal gas ($\beta P = \rho$). This is due to the fact that discretizing the angular variables (like we do it in Z2 and Z4 models), the infinitesimal fluctuations allowed in the original model is not included. Therefore the system of hard needles cannot be examined by *n*-state Zwanzig model at very high pressures. Moreover, the number of orientational states has to exceed two to obtain orientational correlations.

2) V model has allowed us to perform complete analysis, but it can be considered as an approximate model for hard needles. The model gives account of the orientational ordering and the fluctuations even at the high pressure limit. It is proved exactly that the orientational fluctuations gives an extra factor in the ideal gas law such that $\beta P = 2\rho$ at very high densities. Interestingly, the system is only partially ordered even at very high pressures. The divergence of the correlation length indicates that the particles have to rotate together in large blocks at very high densities due to the lack of free room available.

3) The phase behavior of hard needles cannot be described by discrete models and the continuous V model at high pressures. The numerical results with $n$=256 orientational states suggest that the orientational correlation length does not diverge, but it saturates with increasing pressure.

In summary we have presented three analytically solvable models, which are capturing some of the properties of the hard needles. In addition, the V model can be considered as a new hard body model without having definite geometry like the hard Gaussian overlap model [30,35]. The group of exactly solvable one-dimensional models may be extended toward the direction of 3D objects by the same technique. Good start in this direction is the recent study of Schwartz [36] for the system of hard non-spherical beads on a line.

### References

[1] L. Onsager, Ann. N.Y. Acad. Sci. **51**, 627 (1949).

[2] G. J. Vroege and H. N. W. Lekkerkerker, Rep. Prog. Phys. **55** 1241 (1992).

[3] G. Cinacchi G and F. Schmid F, J. Phys.: Condens. Matter **14** 12223 (2002).

[4] H. Hansen-Goos and K. Mecke, Phys. Rev. Lett. **102** 018302 (2009).

[5] J. M. Kosterlitz and D. Thouless, J. Phys. C: Solid State Physics **6**, 1181 (1973).

[6] D. Frenkel and R. Eppenga, Phys. Rev. A **31**, 1776 (1985).

[7] R. L. C. Vink, Eur. Phys. J. B. **72**, 225 (2009).

[8] M. A. Bates and D. Frenkel, J. Chem. Phys. **112**, 10034 (2000).

[9] R. A. Perusquia, J. Peon and J. Quintana, Physica A **345**, 130 (2005).

[10] J. Peon, J. Saucedo-Zugazagoitia, G. Sutmann and J. Quintana-H, J. Chem. Phys. **125**, 104908 (2006).

[11] S. Varga and I. Szalai, J. Mol. Liq. **85**, 11 (2000).

[12] A. Chrzanowska, Acta Physica Polonica B **36**, 3163 (2005).

[13] S. Varga, P. Gurin, J. C. Amos, J. Quintana-H, J. Chem. Phys. **131**, 184901 (2009).

[14] L. Tonks, Phys. Rev. **50,** 955 (1936).

[15] T. T. M. Vo, L.-J. Chen and M. Robert, J. Chem. Phys. **119**, 5607 (2003).

[16] H. Hansen-Goos, C. Lutz, C. Bechinger and R. Roth, Europhysics Letters **74**, 8 (2006).

[17] L. M. Casey and L. K. Runnels, J. Chem. Phys. **51**, 5070 (1969).

[18] J. L. Lebowitz, J. K. Percus and J. Talbot, J. Stat. Phys. **49**, 1221 (1987).

[19] B. Martínez-Haya, J. M. Pastor and J. A. Cuesta, Phys. Rev. E **59**, 1957 (1999).

[20] C. Tutschka, J. Math. Phys. **44**, 5224 (2003).

[21] M. Ginoza and M. Silbert, J. Stat. Phys. **110**, 419 (2003).

[22] M. Moradi and F. Taghizadeh, Physica A. **387**, 6463 (2008).

[23] Y. Kantor and M. Kardar, Phys. Rev. E. **79**, 041109 (2009).

[24] Y. Kantor and M. Kardar, EPL **87** 60002 (2009).

[25] M. H. Ansari, http://arxiv.org/abs/0911.5312.

[26] A. Ghosh and D. Dhar, EPL **78**, 20003 (2007).

[27] M. D. A. Fernandez, D. H. Linares and R. A. J. Pastor EPL, **82**, 50007 (2008).

[28] T. Fischer and R. L. C. Vink EPL, **82**, 56003 (2009).

[29] J. M. Yeomans, Statistical Mechanics of Phase Transitions (Clarendon Press, Oxford, 1992).

[30] B. J. Berne and P. Pechukas, J. Chem. Phys. **56**, 4213 (1972).

[31] B. C. Freasier and L. K. Runnels, J. Chem. Phys. **58**, 2963 (1973).

[32] P. Gurin and S. Varga, Phys. Rev. E. **82**, 041713 (2010).

[33] R. Zwanzig, J. Chem. Phys. **39,** 1714 1175 (1963).

[34] R. W. D. Nickalls, Mathematical Gazette **77**, 354 (1993).

[35] P. Padilla and E. Velasco, J. Chem. Phys. **106**, 10299 (1997).

[36] M. Schwartz, Physica A **389**, 731 (2010).

**Figures**

**Figure 1.** Schematic representation of the quasi-one-dimensional system of hard needles with length *l*. Hard needles rotate freely in a plane, but the centers of the particles are restricted to a line. Overlaps are not allowed between the needles.

**Figure 2.**. (Color online) Equation of state (upper panel) and the corresponding order parameter (lower panel) of four different models: two-state Zwanzig (Z2), four-state Zwanzig (Z4), V and needle (N) fluids. The density and the pressure are dimensionless: $\rho^* = \rho l$ and $P^* = \beta P l$.

**Figure 3.** (Color online) Orientational correlation length as a function of pressure. Curves of the four-state Zwanzig (Z4), V and needle (N) fluids are shown. The correlation length and the pressure are dimensionless: $\xi^* = \xi/l$ and $P^* = \beta P l$.

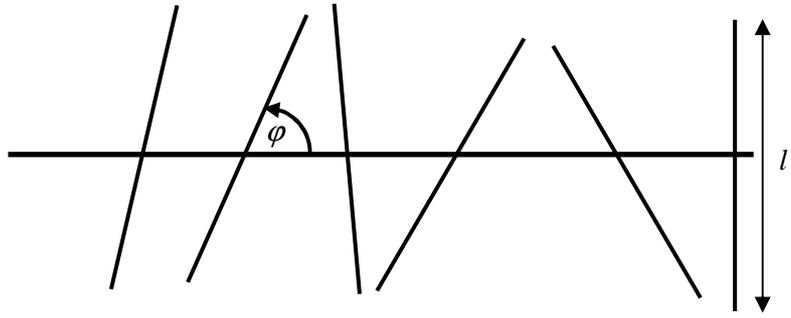

Figure 1.

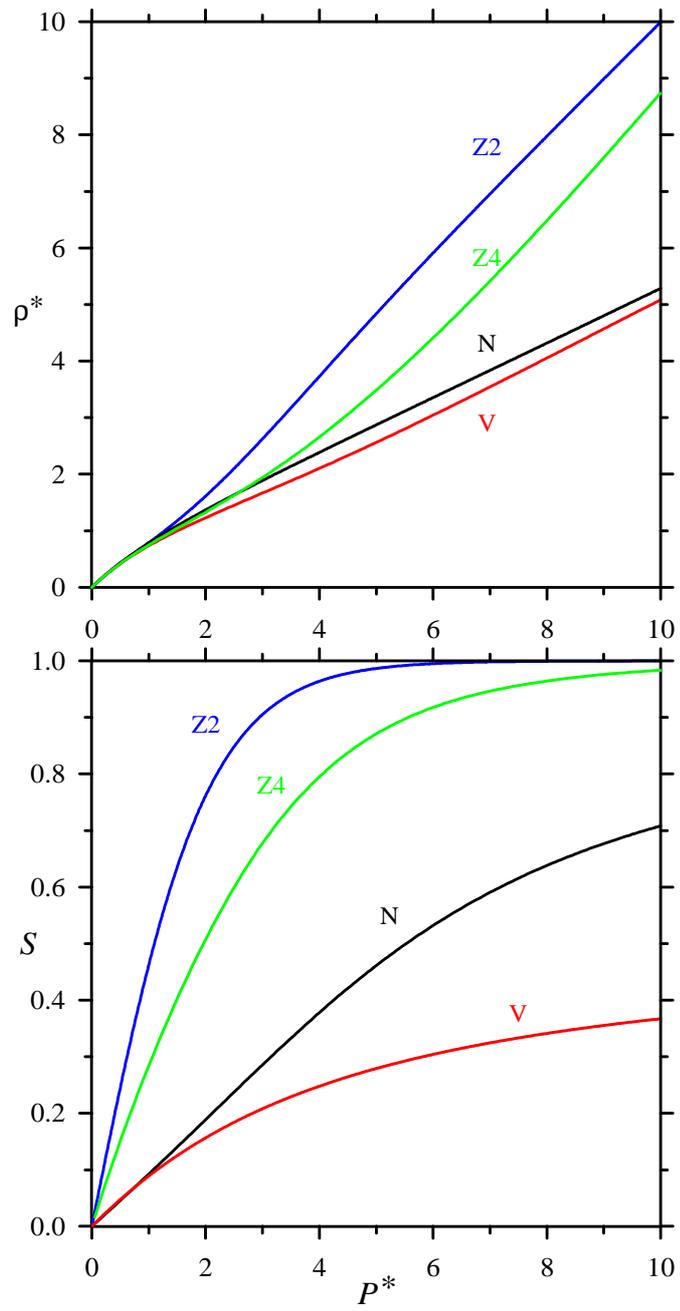

Figure 2.

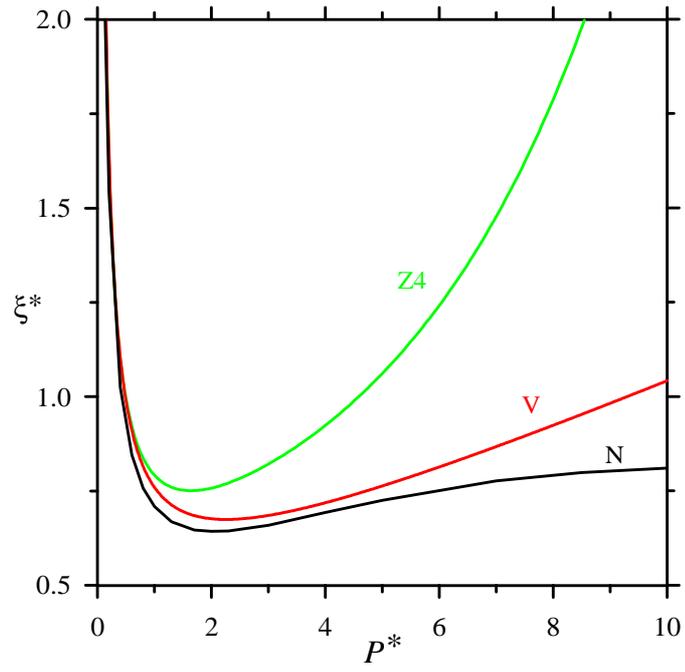

Figure 3.